\documentclass[3p,times,twocolumn]{revtex4}

\usepackage{amssymb}
\usepackage{amsthm}
\usepackage{amsbsy}
\usepackage{amsmath}

\usepackage[figuresright]{rotating}

\begin{document}
\large




\title{Hadron Structure in Chiral Perturbation Theory}


\author {A. Aleksejevs}
\email {aaleksejevs@grenfell.mun.ca}
\affiliation {Grenfell campus of Memorial University, Newfoundland, Canada}

\author {S. Barkanova}
\email {svetlana.barkanova@acadiau.ca}
\affiliation {Acadia University, Nova Scotia, Canada}

\begin{abstract}
We present our predictions for meson form factors for the SU(3)
octet and investigate their impact on the pion electroproduction cross
sections. The electric and magnetic polarizabilities of the SU(3) octet of mesons and
baryons are analyzed in detail. These extensive calculations are made possible
by the recent implementation of semi-automatized calculations in fully-relativistic
chiral perturbation theory, which allows evaluation of polarizabilities
from Compton scattering up to next-to-the-leading order. 
\end{abstract}

\maketitle








\section{Introduction}

The Chiral Perturbation Theory (CHPT) has been tremendously successful
in describing low-energy hadronic properties in the non-perturbative
regime of Quantum Chromodynamics (QCD). Some of the major goals of low-energy QCD are the study of hadronic form factors, which reflect the static structure, and the investigation of the dynamical hadronic response to the external electromagnetic field via
electric and magnetic polarizabilities.
 To study the dependence of the electric and magnetic polarizabilities on the photon
energy, we use the relativistic CHPT
while applying the multipole expansion approach for the Compton structure
functions. Unfortunately, the various versions of CHPT predict a rather
broad spectrum of values for the polarizabilities, introducing theoretical
uncertainty. However, so far CHPT is the only theory available in the
regime of non-perturbative QCD, so our Computational Hadronic
Model (CHM) employed here is based on relativistic CHPT. CHM gives us the opportunity
to avoid the low-energy approximation in the Compton structure functions
and retain all the possible degrees of freedom arising from the SU(3)
chiral Lagrangian. The article is constructed as follows. Section 2 discusses the meson form factor and its role in the
pion electroproduction. Section 3 and 4 are dedicated to the dynamical polarizabilities of the mesons and baryons, respectively. Section 5 briefly summarizes our conclusions.

\section{Pion Form Factor}

The spatial pion electromagnetic form factor has been
addressed in \cite{Ga84,NR,JK,FC,IBG,MT} and is 
under experimental study currently \cite{Fpi}. 
To investigate the behavior of the pion form factor experimentally at  momentum transfers $Q^{2}>0.3\, GeV^{2}$ 
and in the transitional region between long-distance and short-distance QCD, one must use the
charged pion electroproduction process. The two-fold differential
cross section for the exclusive pion electroproduction can be parametrized
by a well known formula \cite{EP1} in terms of photoabsorption
cross sections, where each term corresponds to certain polarization
states of the virtual photon:
\begin{eqnarray}
&&2\pi\frac{d^{2}\sigma}{dtd\phi}=\epsilon\frac{d\sigma_{L}}{dt}+
\frac{d\sigma_{T}}{dt}+ \nonumber \\ \nonumber \\
&&\sqrt{2\epsilon(\epsilon+1)} \frac{d\sigma_{LT}}{dt}\cos\phi 
+\epsilon\frac{d\sigma_{TT}}{dt}\cos2\phi \space . \label{eq:1}
\end{eqnarray}
Here, subscripts L and T correspond to the longitudinal and transverse
polarizations of the virtual photon respectively. Parameter $t$ is
the negative momentum transfer to the hadronic target squared and $\phi$
is the azimuthal angle of the detected hadron in the center-of-mass reference
frame. If the experimental setup has the azimuthal acceptance
\cite{Fpi}, it is possible to determine interference terms, $\sigma_{LT}$
and $\sigma_{TT}$, and then extract the longitudinal term $\sigma_{L}$
by the Rosenbluth separation. In the t-pole approximation, the longitudinal
term, $\sigma_{L}$, is related to the pion form factor, $F{}_{\pi}$,
in the following way:
\begin{eqnarray}
\frac{d\sigma_{L}}{dt}\propto-\frac{tQ^{2}}{t-m_{\pi}^{2}}g_{\pi NN}^{2}(t)F_{\pi}^{2}(Q^{2},t),\label{eq:2} \space ,
\end{eqnarray}
where $g{}_{\pi NN}(t)$ is a pion-proton coupling. Thus, Eq.(\ref{eq:2}) allows the extraction of the pion electromagnetic form factor
for different momentum transfers above the pion production threshold. As it is well known, the determination of the pion form factor from the pion electroproduction is impacted by radiative corrections,  $\delta=\frac{d\sigma_{obs}}{d\sigma_{o}}$.
The radiative corrections to the electron
current and vacuum polarization were calculated in \cite{AAB},
and leading hadronic corrections (two-photon box diagrams) were addressed in \cite{AAB2}. In \cite{AAB2}, it was found that the two-photon box
correction could
reach as much as -20\% for the backward kinematics
($\epsilon\rightarrow0$) and high momentum transfers. 
In order to calculate the form factor of the pion, we use the CHM from \cite{CHM}, which is based on
CHPT. Here, we do one-loop calculations
with a subtractive renormalization scheme for the scale
fixed by the charge radius of the pion: ($<r_{\pi}^{2}>=(0.439\pm0.030)\,\mbox{fm\ensuremath{^{2}}}$).
For the $\pi-\gamma-\pi$ interaction, we arrive at the following renormalized amplitude:
\begin{eqnarray}
M_{r}(q)=e\epsilon\cdot(p'+p) \ \bigg(1-\frac{q^{2}}{f_{\pi}^{2}}f_{1}(q^{2},\Lambda^{2},m_{\pi}^{2})\bigg)-
\nonumber \\
e\frac{\epsilon\cdot q(p'^{2}-p^{2})}{f_{\pi}^{2}}f_{2}(q^{2},\Lambda^{2},m_{\pi}^{2}).\label{eq:3}
\end{eqnarray}
From this, we can form two form factors - one on-shell and another off-shell (if one of the pions is off-shell): 
\begin{eqnarray}
&&F_{on-shell}(q^{2})=1-\frac{q^{2}}{f_{\pi}^{2}}f_{1}(q^{2},\Lambda^{2},m_{\pi}^{2})\label{eq:4}\\
&&F_{off-shell}(q^{2})=-\frac{p'^{2}-p^{2}}{f_{\pi}^{2}}f_{2}(q^{2},\Lambda^{2},m_{\pi}^{2}).\label{eq:5}
\end{eqnarray}
Here, $p$ and $p'$ are momenta of the incoming and outgoing pion, 
 $q$ is the momentum of the virtual photon, and the functions, $f_{1, 2}(q^{2},Q^{2},m_{\pi}^{2})$, 
depend on one- and two-point Passarino-Veltman functions. To incorporate CHPT
into the calculations of the two-photon box correction,
we fit the pion on-shell monopole formfactor to the form
factor in Eq.\ref{eq:4} for the $Q^{2}<0.3\,\mbox{GeV}^{2}$ and get $F_{on-shell}(Q^{2})=\frac{\Phi^{2}}{\Phi^{2}+Q^{2}}$
with $\Phi=0.73\,\mbox{GeV}$. A reason to choose the monopole form
factor is its asymptotic behavior at high momentum transfers, $F_{on-shell}(Q^{2})\bigg|_{Q^{2}\rightarrow\infty}=\frac{8\pi\alpha_{s}(Q^{2})f_{\pi}^{2}}{Q^{2}}$, driven by the perturbative QCD.
For the off-shell form factor in Eq.(\ref{eq:5}), we also choose the monopole
form, which is fitted to the CHPT off-shell form factor, so
we get $F_{off-shell}(Q^{2},t)=(m_{\pi}^{2}-t)\frac{Q^{2}}{\Omega^{2}+Q^{2}}$
with $\Omega=2.5\,\mbox{GeV.}$

\begin{figure}[!htpb]
\begin{centering}
\includegraphics[scale=0.28]{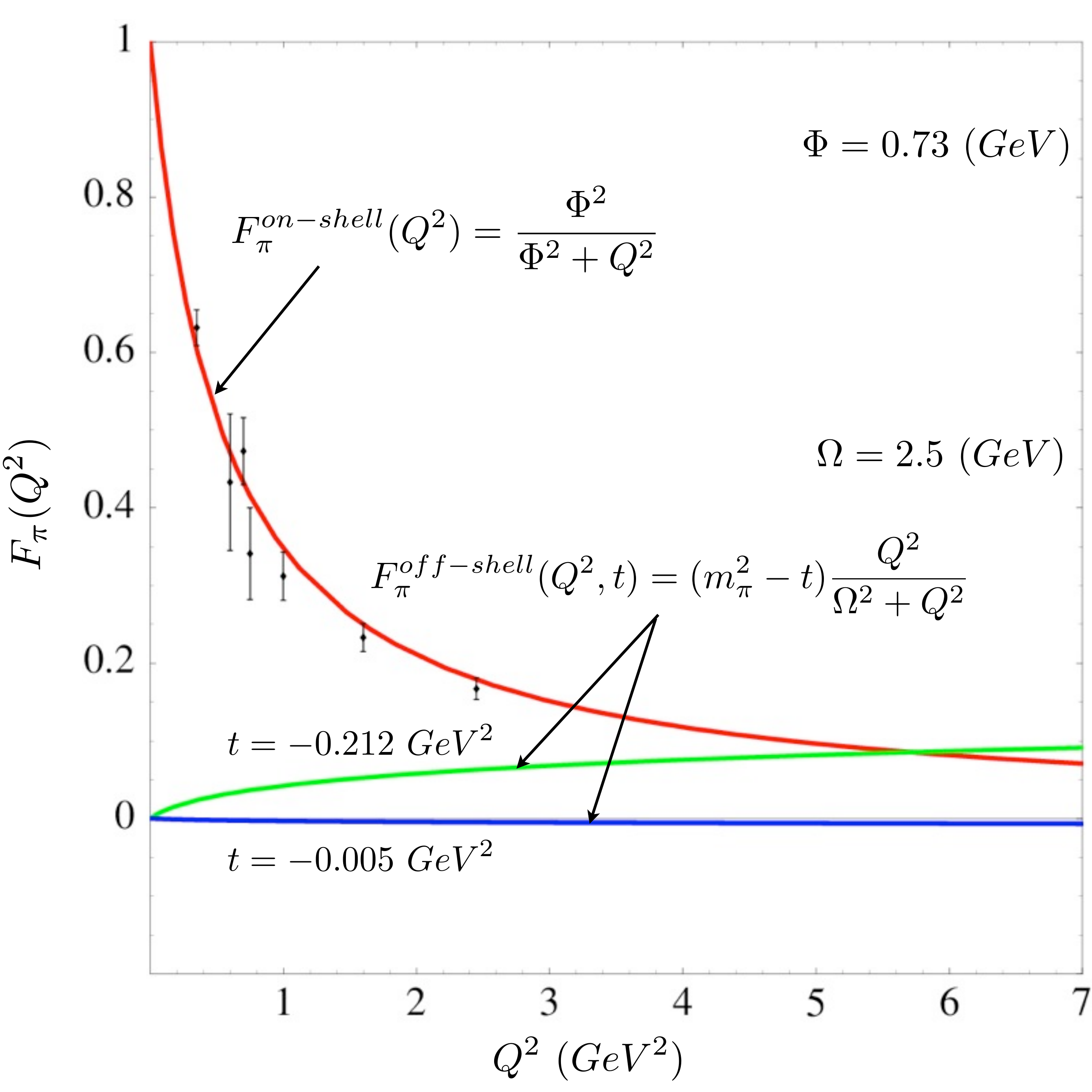}
\par\end{centering}

\caption{On-shell and off-shell pion form factors. The red line shows the monopole
fit of the on-shell formfactor to CHPT results. Experimental data
are taken from \cite{Amend,Acker,Brauel,Hub}. The off-shell form factor
is shown for both low ($t=-0.005\,\mbox{GeV}^{2}$, blue line) and
high ($t=-0.212\,\mbox{GeV}^{2}$, green line) hadronic momentum transfers.}

\label{fig3}
\end{figure}

\begin{figure*}[!htpb]
\begin{centering}
\includegraphics[scale=0.28]{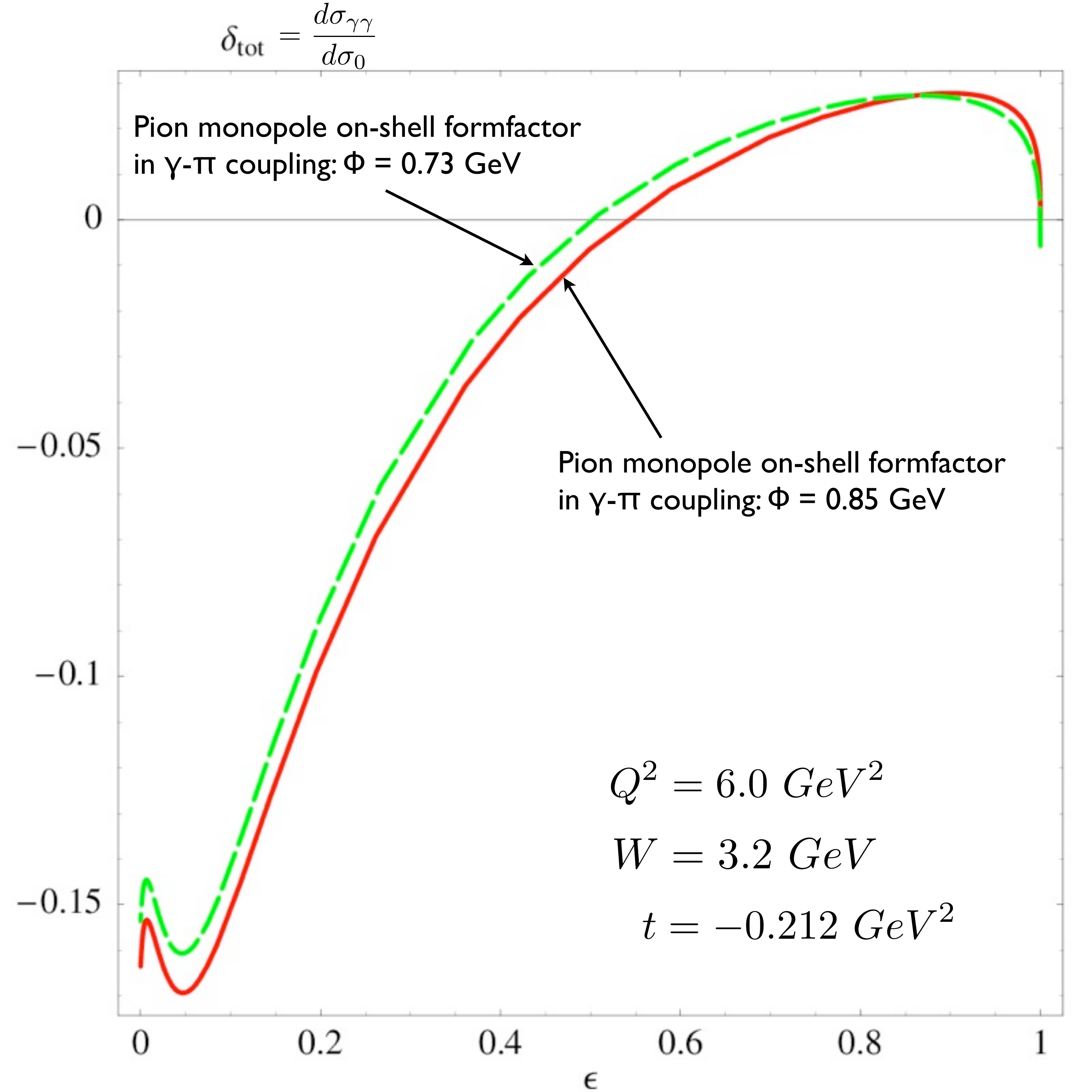} \includegraphics[scale=0.28]{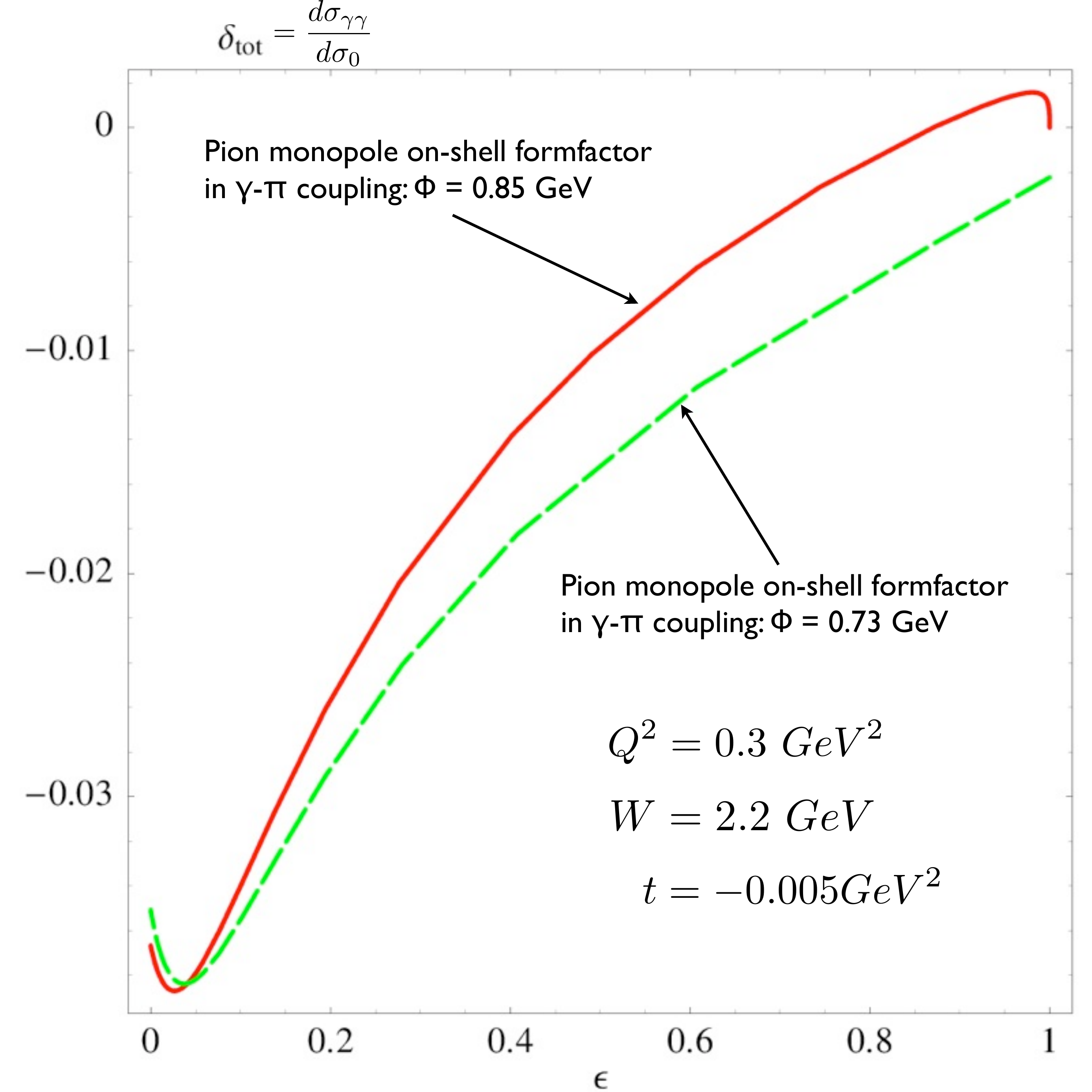}
\par\end{centering}

\caption{Two-photon box radiative correction for the exclusive pion electroproduction
for $Q^{2}=6.0\,\mbox{GeV}^{2}$ (left plot) and $Q^{2}=0.3\,\mbox{GeV}^{2}$
(right plot). The corrections calculated with the monopole pion form factor
for $\Phi=0.73\,\mbox{GeV}$ and $\Phi=0.85\,\mbox{GeV}$ are shown
by the dashed green and solid red lines, respectively.}

\label{fig3-1}
\end{figure*}

From Fig.\ref{fig3}, one can see
that our fit is in rather good agreement with experimental data.
We now use the fitted form factors in the calculations of
the two-photon box pion electroproduction correction with the same
tools for the exact calculations as in \cite{AAB2}, for high ($Q^{2}=6.0\,\mbox{GeV}^{2}$)
and low ($Q^{2}=0.3\,\mbox{GeV}^{2}$) momentum transfers (see Fig.\ref{fig3-1}).

Two lines, $\Phi=0.73\,\mbox{GeV}$ (green dashed line) and $\Phi=0.85\,\mbox{GeV}$
(red solid line), illustrate the box correction sensitivity to the choice of the scale in the form factor. 
As can be clearly seen from the right plot of Fig.\ref{fig3-1}, for the low momentum transfer
the correction is rather sensitive to
the scale, which induces an additional
theoretical uncertainty due to the box correction model dependence. For the high momentum transfer
(left plot of Fig.\ref{fig3-1}, we observe that the correction
does not change much with the change of the scale in the form factor. Thus, we observe a  reduced degree of model dependence
in the correction for high momentum transfers. For the off-shell
part of the pion form factor, we have observed a contribution of less than 1\% 
to the correction, which certainly diminishes any role of the off-shell
form factor in the pion electroproduction process. 

\section{Dynamical Polarizabilities of Mesons }

Experimentally, a unique opportunity to study the dynamical structure of hadrons over
a wide kinematic range is provided by Compton scattering. In the non-relativistic approximation, the Hamiltonian related to
the meson internal structure can be represented
as 
\begin{eqnarray}
H_{eff}=-\frac{1}{2}4\pi\alpha_{E}\vec{E}^{2}-\frac{1}{2}4\pi\beta_{M}\vec{H}^{2},\label{eq:6}
\end{eqnarray}
where $\alpha_{E}$ and $\beta_{M}$ are electric and magnetic
polarizabilities of the meson, respectively. Up to now,
only charged and neutral pion polarizabilities have been measured,
producing a broad range of values \cite{Ahrens,Antipov,Ba92}. Precision measurements of the charged pion polarizability through the
Primakoff two-pion photo-production process with linearly polarized
photons is planned at JLab. The cross section for this process,
\begin{eqnarray}
\frac{d^{2}\sigma}{d\Omega_{\pi\pi}dW_{\pi\pi}}=\frac{2\alpha_{f}Z^{2}}{\pi^{2}}\frac{E_{\gamma}^{4}\beta^{2}}{W_{\pi\pi}}\frac{\sin\theta_{\pi\pi}^{2}}{Q^{4}}|F(Q^{2})|^{2}
\nonumber \\ \nonumber \\
\sigma(\gamma\gamma\rightarrow\pi\pi)(1+2P_{\gamma}\cos2\phi_{\pi\pi}),\label{eq:7}
\end{eqnarray}
is related to the photon fusion cross section, $\sigma(\gamma\gamma\rightarrow\pi\pi)$,
which can be easily turned into a Compton cross section by means of crossing symmetry. In Eq.(\ref{eq:7}), $W_{\pi\pi}$ is the $\pi\pi$
invariant mass, Z is the atomic number, $E_{\gamma}$ is the energy
of the incident photon, $F(Q)$ is the electromagnetic form factor
for the proton target, $\theta_{\pi\pi}$ is the lab angle for the
two pions, $P_{\gamma}$ is the incident photon polarization and $\phi_{\pi\pi}$
is the azimuthal angle of the $\pi\pi$ system. One can
relate photon fusion cross section, $\sigma(\gamma\gamma\rightarrow\pi\pi)$,
to the polarizabilities of the pion \cite{Ba92,Do93} as
\begin{eqnarray}
\sigma_{\gamma\gamma\rightarrow\pi\pi}(|\cos\theta|<Z)=\frac{\kappa}{256\pi s^{2}}\int_{t_{a}}^{t_{b}}dt
\nonumber \\ \nonumber \\ \nonumber \\
\Bigg(\bigg|m_{\pi}^{2}B_{o}-8\pi sm_{\pi}\beta+\frac{4\pi}{m_{\pi}}(\alpha+\beta)st\bigg|^{2}+\label{eq:8} \nonumber \\
\nonumber \\ \nonumber \\
\bigg|B_{o}+\frac{4\pi s}{m_{\pi}}(\alpha+\beta)\bigg|^{2}\frac{(m_{\pi}^{4}-tu)^{2}}{s^{2}}\Bigg),
\end{eqnarray}
where 
\begin{eqnarray*}
  &\displaystyle{B_{o}=16\pi\alpha_{f}{\displaystyle \frac{s}{(t-m_{\pi}^{2})(u-m_{\pi}^{2})}}|q|}, 
\end{eqnarray*}
and
\begin{eqnarray*}
  &\displaystyle t_{b,a}=m_{\pi}^{2}-\frac{1}{2}s\pm\frac{sZ}{2}\beta(s).
\end{eqnarray*}
Here, $s,\, t$ and $u$ are the Mandelstam variables, $|q|$ is
the meson charge, $\beta(s)=\sqrt{\frac{s-4m_{\pi}^{2}}{s}}$ is
the center-of-mass velocity of produced pions and $\kappa=1$ or $2$  for a
neutral or charged pion, respectively.
Eqs.\ref{eq:7} and \ref{eq:8} are valid for any meson, not just pions. For real Compton scattering, we can construct an
invariant amplitude,
\begin{eqnarray}
\ M(\gamma\pi\rightarrow\gamma'\pi)=\epsilon'^{\mu}\epsilon^{\nu}\ M_{\mu\nu}.\label{eq:9}
\end{eqnarray}
Here $\boldsymbol{\epsilon}\:\mbox{and }\boldsymbol{\epsilon'}$ are
polarization vectors of incoming and outgoing photons, respectively,
and $M_{\mu\nu}$ is the Compton tensor, which is related to two Compton
structure functions ($A(s,t)$ and $B(s,t)$) in the following way:
\begin{eqnarray}
M_{\mu\nu}=A(s,t)\ T_{\mu\nu}^{(1)}+B(s,t)\ T_{\mu\nu}^{(2)}.\label{eq:10}
\end{eqnarray}
The Lorentz tensors, $T_{\mu\nu}^{(1)}$ and $T_{_{\mu\nu}}^{(2)}$, are
\begin{eqnarray}
&T_{\mu\nu}^{(1)}  =-\displaystyle{\frac{t}{2}g_{\mu\nu}-k_{3,\mu}k_{1,\nu}}\nonumber \\
\nonumber \\
T_{\mu\nu}^{(2)}& =\displaystyle{\frac{1}{2t}(s-m_{\pi}^{2})(u-m_{\pi}^{2})g_{\mu\nu}+k_{2,\mu}k_{2,\nu}+}\nonumber \\ 
\nonumber \\
&\displaystyle{\frac{s-m_{\pi}^{2}}{t}k_{3,\mu}k_{3,\nu}-\frac{u-m_{\pi}^{2}}{t}k_{2,\mu}k_{1,\nu}}.\label{eq:11} 
\end{eqnarray}
The amplitude related
to the meson electric and magnetic polarizabilities can now be rewritten as
\begin{eqnarray}
\ M(\gamma\pi\rightarrow\gamma'\pi)=\alpha\omega^{2}(\boldsymbol{\epsilon}{}^{\prime*}\cdot\boldsymbol{\epsilon})+\beta\omega^{2}({\bf s}{}^{\prime*}\cdot{\bf s}),\label{eq:12}
\end{eqnarray}
where $\omega$ is the photon energy and $\boldsymbol{s}=(\boldsymbol{k}\times\boldsymbol{\epsilon})$
denotes the magnetic vector. Combining Eq.(\ref{eq:10}) and (\ref{eq:12}),
we get a connection between the polarizabilities and Compton structure functions:
\begin{eqnarray}
\alpha(s,t)=-\frac{1}{8\pi m}\bigg(A(s,t)+\frac{s-3m^{2}}{t}B(s,t)\bigg)\nonumber \\
\label{eq:13}\\
\beta(s,t)=\frac{1}{8\pi m}\bigg(A(s,t)+\frac{s-m^{2}}{t}B(s,t)\bigg).\nonumber 
\end{eqnarray}
Terms in Eq.(\ref{eq:13}) are  clearly energy-dependent, so we call them dynamical. In the limit when $s\rightarrow m^{2}$
and $t\rightarrow0$, we recover the static values of the polarizabilities.
Using CHM and restricting our calculations to one-loop in CHPT (two-loop
SU(2) calculations were done in \cite{Ga06} and showed a rather
small impact on the cross section in Primakoff reaction), we calculate
the dynamical polarizabilities of mesons including the entire SU(3) octet
of mesons in the loop integrals. 

\begin{figure*}[!htpb]
\begin{centering}
\includegraphics[scale=0.28]{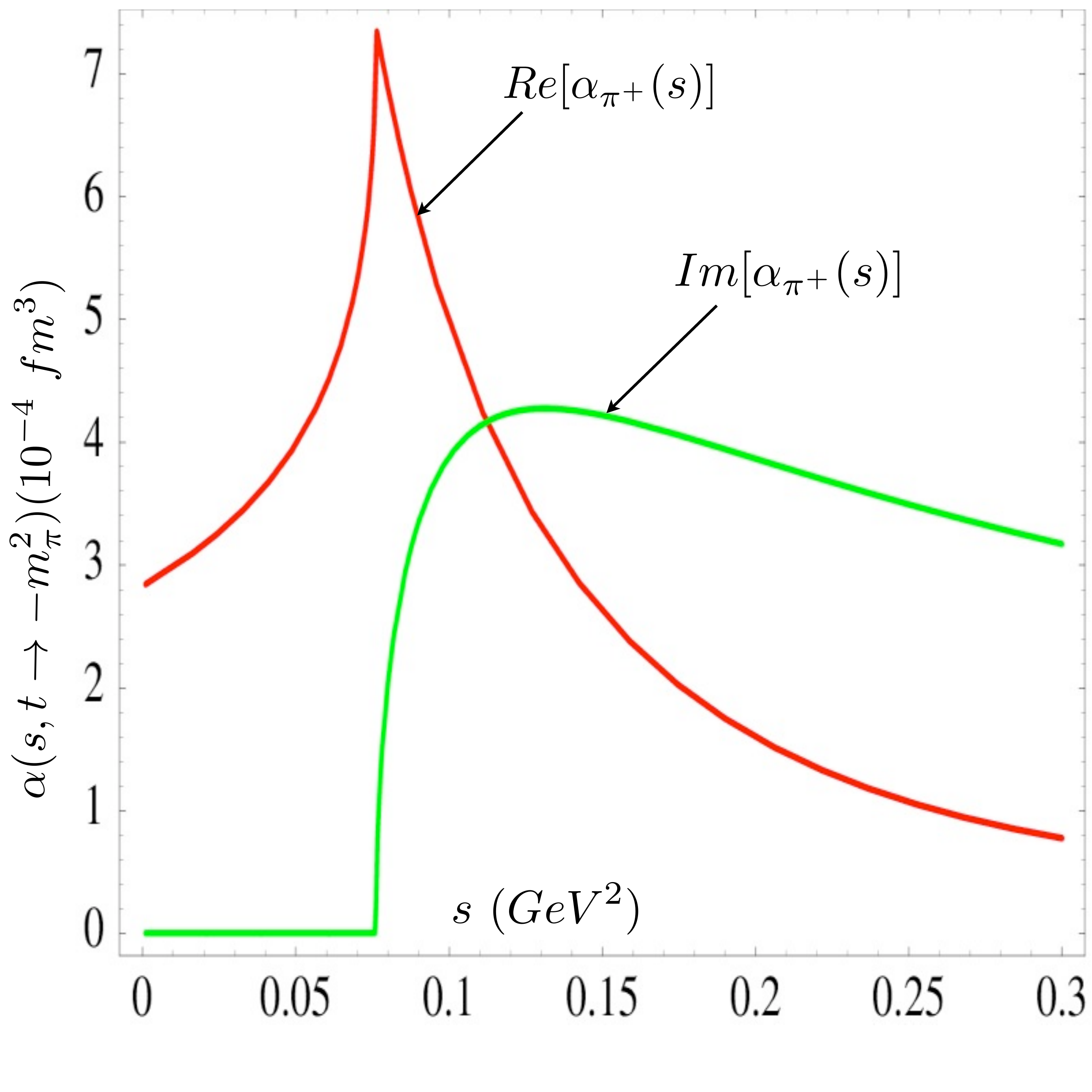} \includegraphics[scale=0.29]{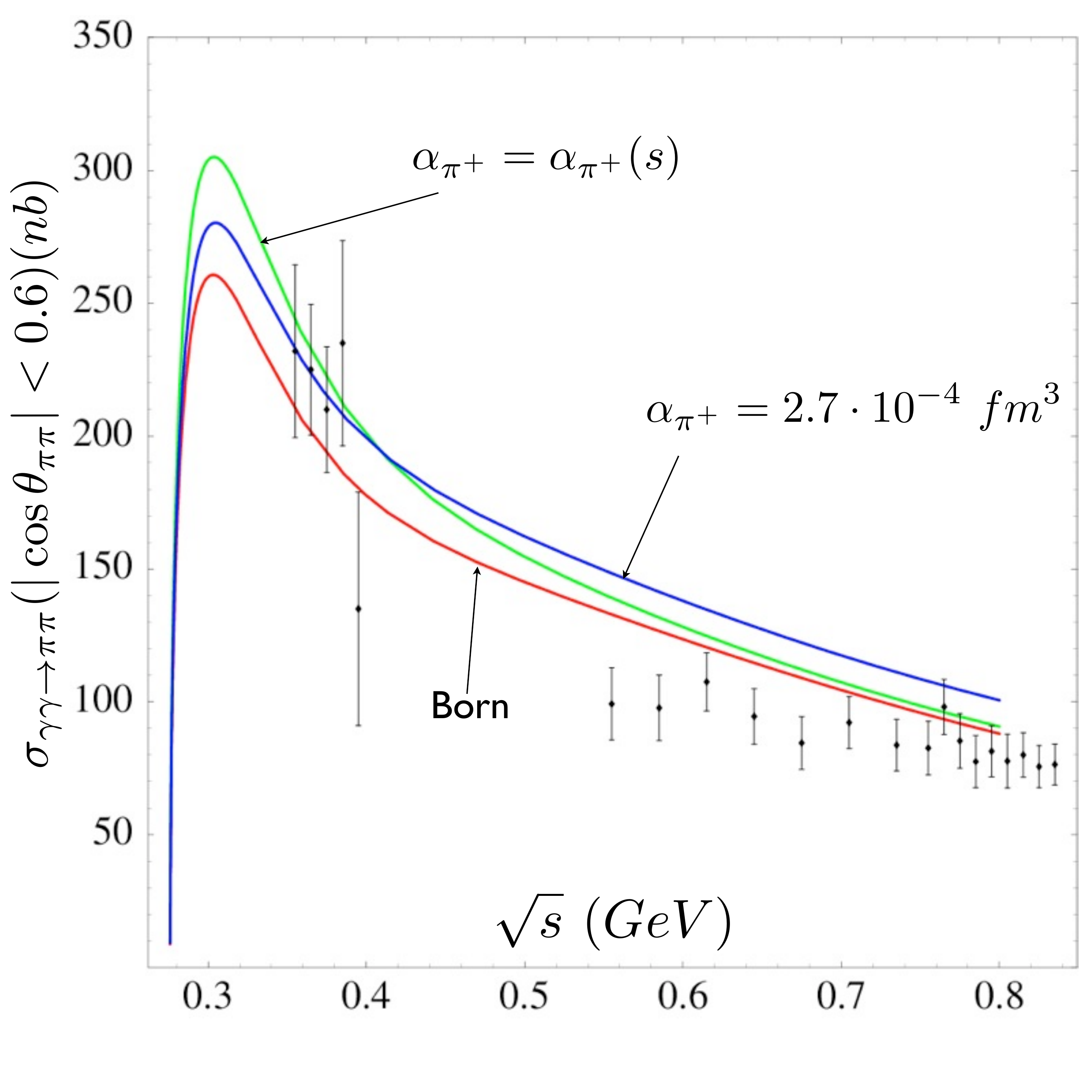}
\par\end{centering}

\caption{Left graph: energy dependence of the pion electric polarizabilty. Right graph: 
$\gamma\gamma\rightarrow\pi\pi$ cross section, where red line corresponds
to the Born cross section, blue line is the cross section related
to the static CHPT value of polarizability and green line shows the cross section with dynamical polarizabilities from Eq.(\ref{eq:13}).
The data points are taken from MARK-II \cite{Bo92}.}
\label{fig5}
\end{figure*}

In addition, we include a structure-dependent pole contribution arising from the
vector mesons in similar fashion to \cite{Ba92,Do93}. We now have the following static electric and
magnetic polarizabilities (in units of $10^{-4}\mbox{fm}^{3}$):
\begin{eqnarray}
&&\alpha_{\pi^{\pm}}=\frac{8\alpha_{f}}{m_{\pi}f_{\pi}^{2}}(L_{9}+L_{10})=2.83; \nonumber \\ 
&&\beta_{\pi^{\pm}}=-\alpha_{\pi^{\pm}}+\frac{m_{\pi}}{4\pi}\frac{G_{\rho}}{M_{\rho}^{2}-m_{\pi}^{2}}=-2.76\nonumber \\
\label{eq:14} \nonumber \\ \nonumber \\
&&\alpha_{\pi^{0}}=-\frac{\alpha_{f}}{48\pi^{2}m_{\pi}f_{\pi}^{2}}=-0.50; \nonumber \\
&&\beta_{\pi^{o}}=-\alpha_{\pi^{o}}+\frac{m_{\pi}}{4\pi}\sum_{V=\rho,\omega}\frac{G_{V}}{M_{V}^{2}-m_{\pi}^{2}}=1.25 \space .
\end{eqnarray}
The pion coupling constant is $f_{\pi}=\sqrt{2}F_{\pi}=130.7\,\mbox{MeV}$,
$\alpha_{f}$ is the fine structure constant, the low energy constants are $L_{9}=(5.99\pm0.43)\cdot10^{-3}$
and $L_{10}=(-4.5\pm0.7)\cdot10^{-3}$, and for the vector mesons coupling
constants we use $G_{\rho}=0.044\,\mbox{GeV}$ and $G_{\omega}=0.495\,\mbox{GeV}$.
The polarizabilities
for the octet of mesons are summarized in Table \ref{tbl1}.
\begin{table}
\begin{centering}
\begin{tabular}{|c|c|c|}
\hline 
($10^{-4}\,\mbox{fm}^{3}$) & $\alpha-\beta$ & $\alpha+\beta$\tabularnewline
\hline 
\hline 
$\pi^{\pm}$ & 5.59 & 0.07\tabularnewline
\hline 
$\pi^{0}$ & -1.75 & 0.75\tabularnewline
\hline 
$\eta$ & -0.044 & 0.0\tabularnewline
\hline 
$K^{\pm}$ & 0.88 & 0.0\tabularnewline
\hline 
$K^{0}$ & 0.0032 & 0.0\tabularnewline
\hline 
\end{tabular}
\par\end{centering}

\caption{Meson electric and magnetic static polarizabilities}

\label{tbl1}
\end{table}

\begin{figure*}[!htpb]
\begin{centering}
\includegraphics[scale=0.28]{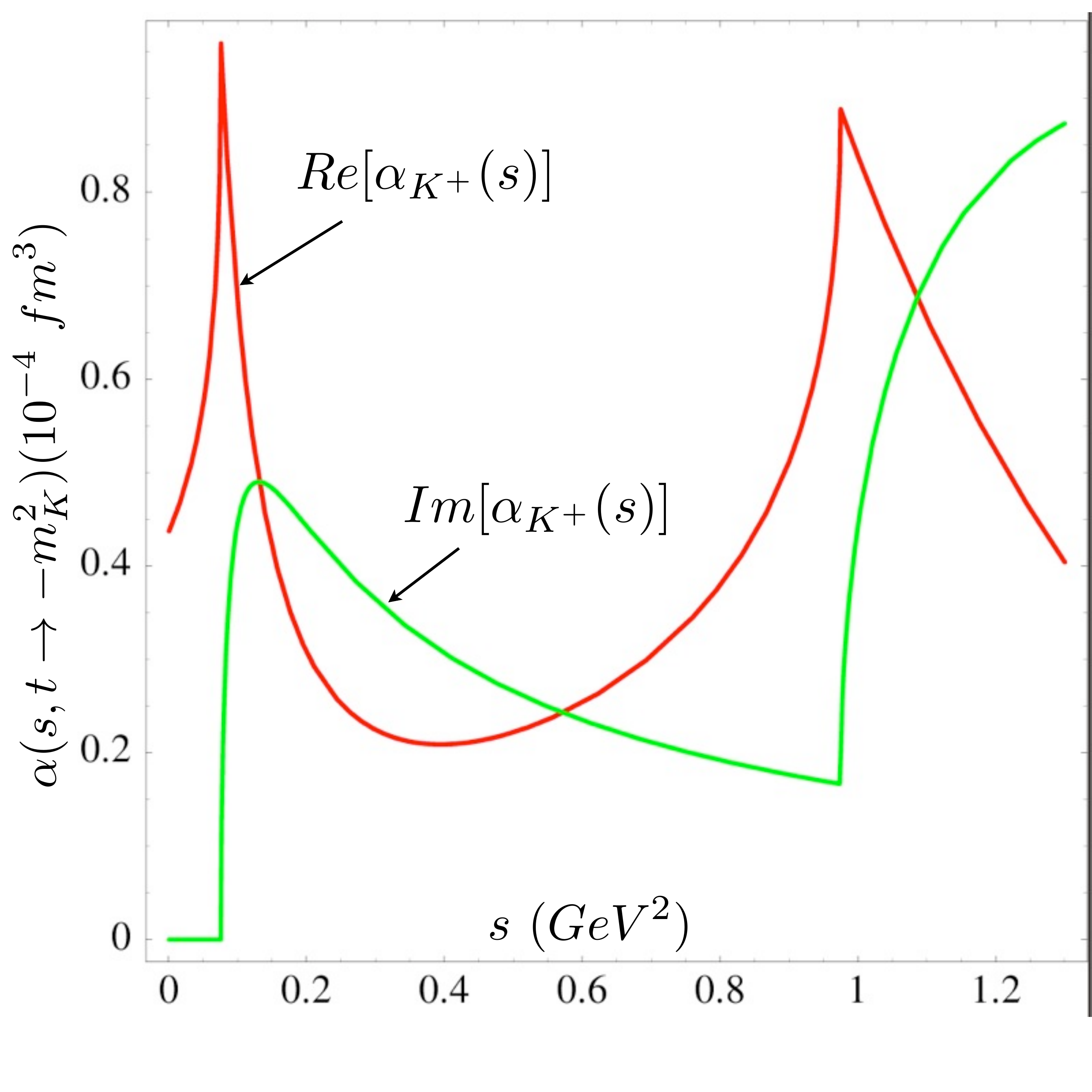} \includegraphics[scale=0.28]{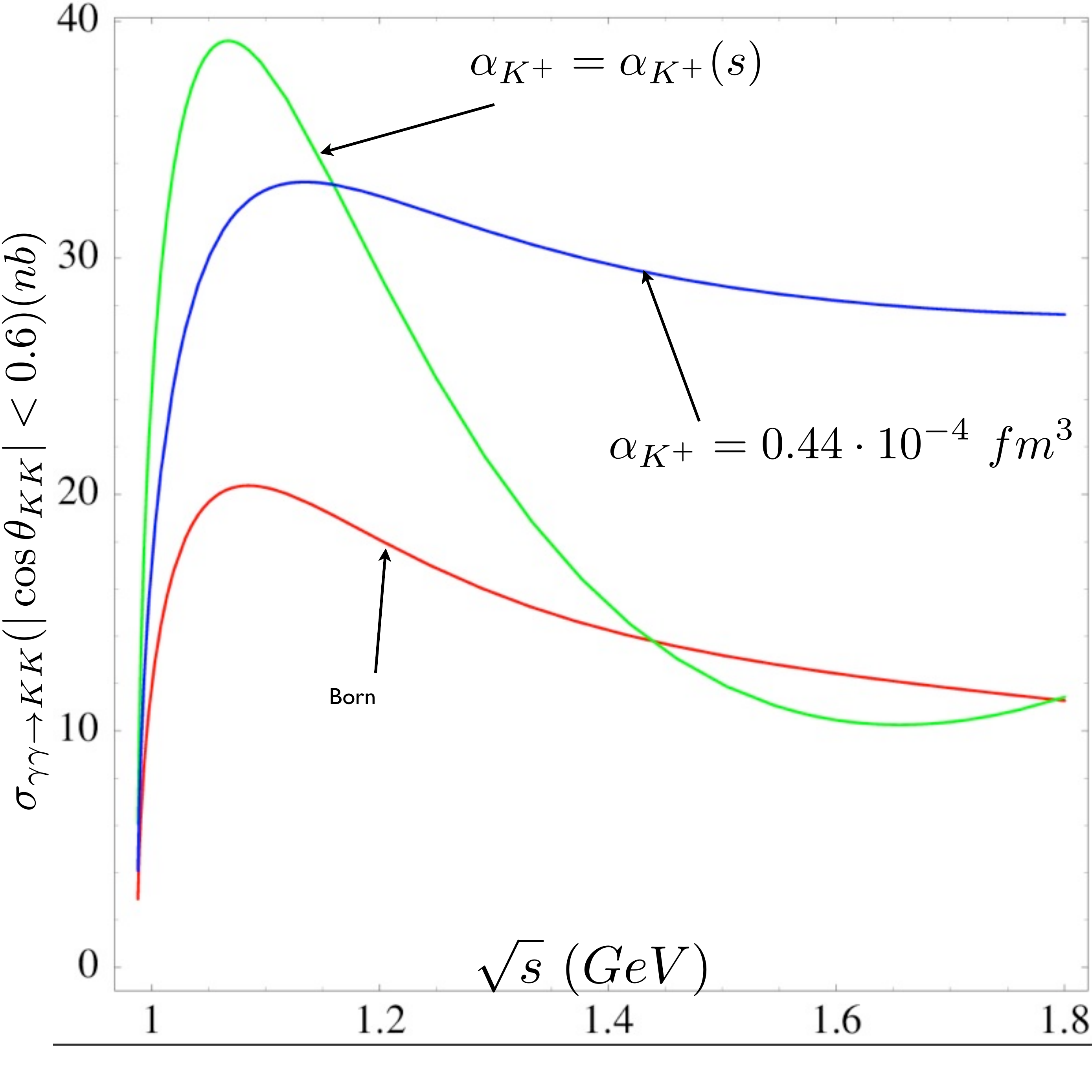}
\par\end{centering}

\caption{Charged kaon dynamical polarizability and prediction for the $\gamma\gamma\rightarrow K^{+}K^{-}$
cross section. On the left plot, red line is the Born cross section,
blue line corresponds to the static polarizability and green line
to the dynamical polarizability cross section.}

\label{fig6}
\end{figure*}

To extract the static pion polarizabilities from the $\gamma\gamma\rightarrow\pi\pi$
cross section, one can use Eq.(\ref{eq:13}) (with crossing symmetry $s\rightarrow t$ and $t\rightarrow s$)
substituted into Eq.(\ref{eq:8}). The important role of the energy dependence of the polarizabilities
in the description of the $\gamma\gamma\rightarrow\pi\pi$ cross section can be seen in Fig.\ref{fig5}.
In Fig.\ref{fig6}, it is clearly visible that the dynamical polarizability
of the kaon has a very strong influence on the $\gamma\gamma\rightarrow K^{+}K^{-}$
cross section. 

\section{Dynamical Polarizabilities of Baryons}

The electric and magnetic polarizabilities of baryons introduce an
additional contribution into the effective Hamiltonian for baryons in the
electromagnetic field in the same way as defined in Eq.(6). 
The current PDG \cite{PDG} experimental values for the
electric and magnetic polarizabilities for proton and neutron are (in units of $10^{-4} \mbox{fm}^3$):
\begin{eqnarray*}
\alpha_{p}=12.0\pm0.6; &  & \beta_{p}=1.9\pm0.5;\\
\alpha_{n}=11.6\pm1.5; &  & \beta_{n}=3.7\pm2.0 \space .
\end{eqnarray*}
In order to evaluate polarizabilities theoretically, one can use Compton
scattering and relate an amplitude to the set of Compton structure
functions $R_{i}$ \cite{Babusci} in the following way:
\begin{eqnarray}
&\displaystyle{\frac{1}{8\pi W}}M(\gamma B\rightarrow\gamma'B) =  R_{1}(\boldsymbol{\epsilon}{}^{\prime*}\cdot\boldsymbol{\epsilon})+R_{2}({\bf s}{}^{\prime*}\cdot{\bf s})+ \nonumber \\
&iR_{3}\boldsymbol{\sigma}\cdot(\boldsymbol{\epsilon}{}^{\prime*}\times\boldsymbol{\epsilon})+iR_{4}\boldsymbol{\sigma}\cdot({\bf s}{}^{\prime*}\times{\bf s})+\nonumber \\
&iR_{5}((\boldsymbol{\sigma}\cdot\hat{{\bf k}})({\bf s^{\prime*}}\cdot\boldsymbol{\epsilon})-(\boldsymbol{\sigma}\cdot\hat{{\bf k}}^{\prime})({\bf s}\cdot\boldsymbol{\epsilon^{\prime*}}))+\nonumber \\
&iR_{6}((\boldsymbol{\sigma}\cdot\hat{{\bf k}}^{\prime})({\bf {\bf s^{\prime*}}}\cdot\boldsymbol{\epsilon})-(\boldsymbol{\sigma}\cdot\hat{{\bf k}})({\bf s}\cdot\boldsymbol{\epsilon^{\prime*}})) \space . \label{eq:4-1}
\end{eqnarray}
Here, $W=\omega+\sqrt{\omega^{2}+m_{B}^{2}}$ is the center of mass
energy and $\omega$ is the energy of the incoming photon. The unit magnetic
vector (${\bf s}=(\hat{{\bf k}}\times\boldsymbol{\epsilon}$)), the polarization
vector ($\boldsymbol{\epsilon}$) and the unit momentum of the photon
(${\displaystyle \hat{{\bf k}}=\frac{{\bf k}}{k}}$) are denoted with
a prime in the case of the outgoing photon. Although the choice
of the basis for the invariant Compton amplitude is not unique and
can be defined differently \cite{Babusci}, the basis in Eq.(\ref{eq:4-1}) is more convenient for the evaluation of the polarizabilities
because in this basis the structure functions, $R_{i}$, are directly related to the
electric, magnetic and spin-dependent polarizabilities in the multipole
expansion. 

It is well known that parameters such as polarizabilities can be determined by 
the non-Born contributions to the Compton structure functions. This includes loops (up to
the given order of perturbation) and structure-dependent pole contributions,
such as tree-level baryon resonance excitations and the WZW anomalous
interaction (contributing only into backward spin-dependent polarizability).
If in the multipole expansion of the Compton structure functions \cite{multi-1,multi-2,multi-3}
we keep only the dipole-dipole and dipole-quadrupole transitions,
we can obtain simple equations connecting non-Born (NB) structure
functions to the polarizabilities of the baryon: 
\begin{eqnarray}
&R_{1}^{NB}=\omega^{2}\alpha_{E1};\ \ R_{2}^{NB}=\omega^{2}\beta_{M1}; \nonumber \\ \nonumber \\
&R_{3}^{NB}=\omega^{3}(-\gamma_{E1E1}+\gamma_{E1M2}); \nonumber \\ \nonumber \\
&R_{4}^{NB}=\omega^{3}(-\gamma_{M1M1}+\gamma_{M1E2});\nonumber \\ \nonumber \\
&R_{5}^{NB}=-\omega^{3}\gamma_{M1E2};\ \ R_{5}^{NB}=-\omega^{3}\gamma_{E1M2}.\label{eq:6-1}
\end{eqnarray}
Although the polarizabilities used in the Eq.(\ref{eq:6-1}) are defined
as constants, it is essential to treat them as energy-dependent quantities
\cite{Griesshammer} because the
Compton scattering experiments were performed with 50 to 800 MeV photons
and hence require additional theoretical information to extrapolate
the results to zero-energy parameters. The
polarizabilities become energy-dependent due to the internal relaxation
mechanisms, resonances, and particle production thresholds. Accordingly,
if for the static polarizabilities we only keep order up to $\mathcal{O}(\omega^{2})$
for $R_{1,2}$ and up to $\mathcal{O}(\omega^{3})$ for $R_{3,4,5,6}$ then
for energy-dependent dynamical polarizabilities we keep all orders
in $\omega$ in the Compton structure functions. The calculation of the Compton structure
functions up to the one-loop order in the framework of relativistic
CHPT was made possible by CHM \cite{CHM}.
In addition, the structure-dependent pole contribution to the nucleon
polarizabilities has been taken into account in the form of the nucleon
$\Delta$-resonance excitation. A Lagrangian which describes nucleon-to-resonance
radiative transition is given in the form of a contact term:
\begin{eqnarray}
 & \displaystyle{\mathcal{L}^{\Delta N\gamma}=i\Theta\frac{e}{\Lambda}\bar{N}\gamma^{\mu}\gamma_{5}Q\Delta^{\nu}F_{\mu\nu},}\nonumber \\
\label{eq:9-1}\\
 & F_{\mu\nu}=\partial_{\mu}\mathcal{A}_{\nu}-\partial_{\nu}\mathcal{A}_{\mu}.\nonumber 
\end{eqnarray}
Here, $\Lambda\sim1\, GeV$ is the scale of chiral symmetry breaking
and $\Theta$ is the coupling strength, determined from the branching
ratio of the radiative decay, $\Delta\rightarrow N\gamma$.
The $\Delta$ resonance propagator is described by the propagator
of the 3/2 spin Rarita-Schwinger field:
\begin{eqnarray}
\Pi_{\Delta}^{\mu\nu}=\frac{1}{2m}\frac{\not p+m}{p^{2}-m^{2}+im\Gamma}\Big(g^{\mu\nu}-\frac{1}{3}\gamma^{\mu}\gamma^{\nu}-\nonumber \\ \nonumber \\
\frac{2p^{\mu}p^{\nu}}{3m^{2}}+\frac{p^{\mu}\gamma^{\nu}-p^{\nu}\gamma^{\mu}}{3m}\Big).\label{eq:10-1}
\end{eqnarray}

\begin{figure*}[!htpb]
\begin{centering}
\begin{tabular}{cc}
\includegraphics[scale=0.37]{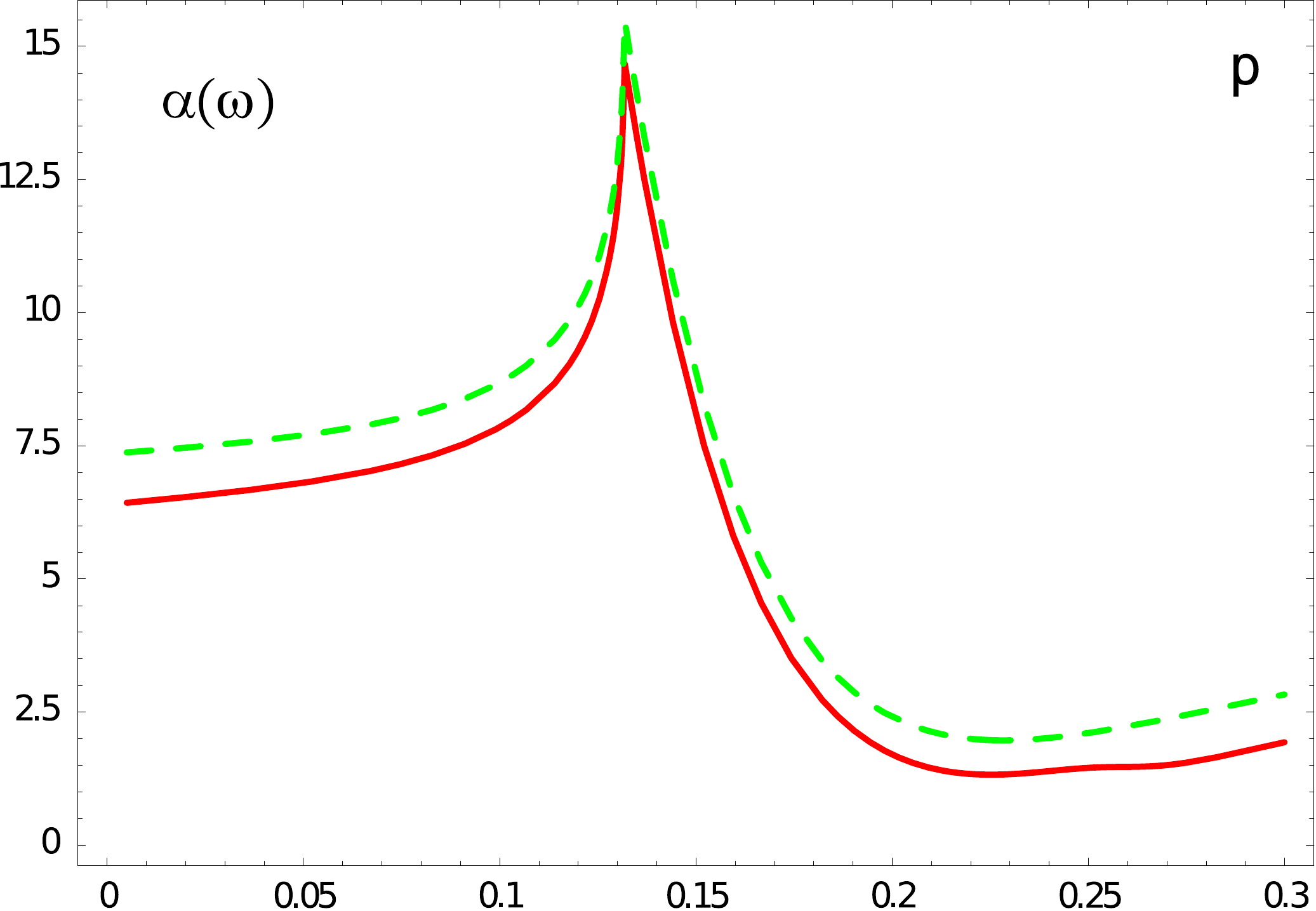} & \includegraphics[scale=0.37]{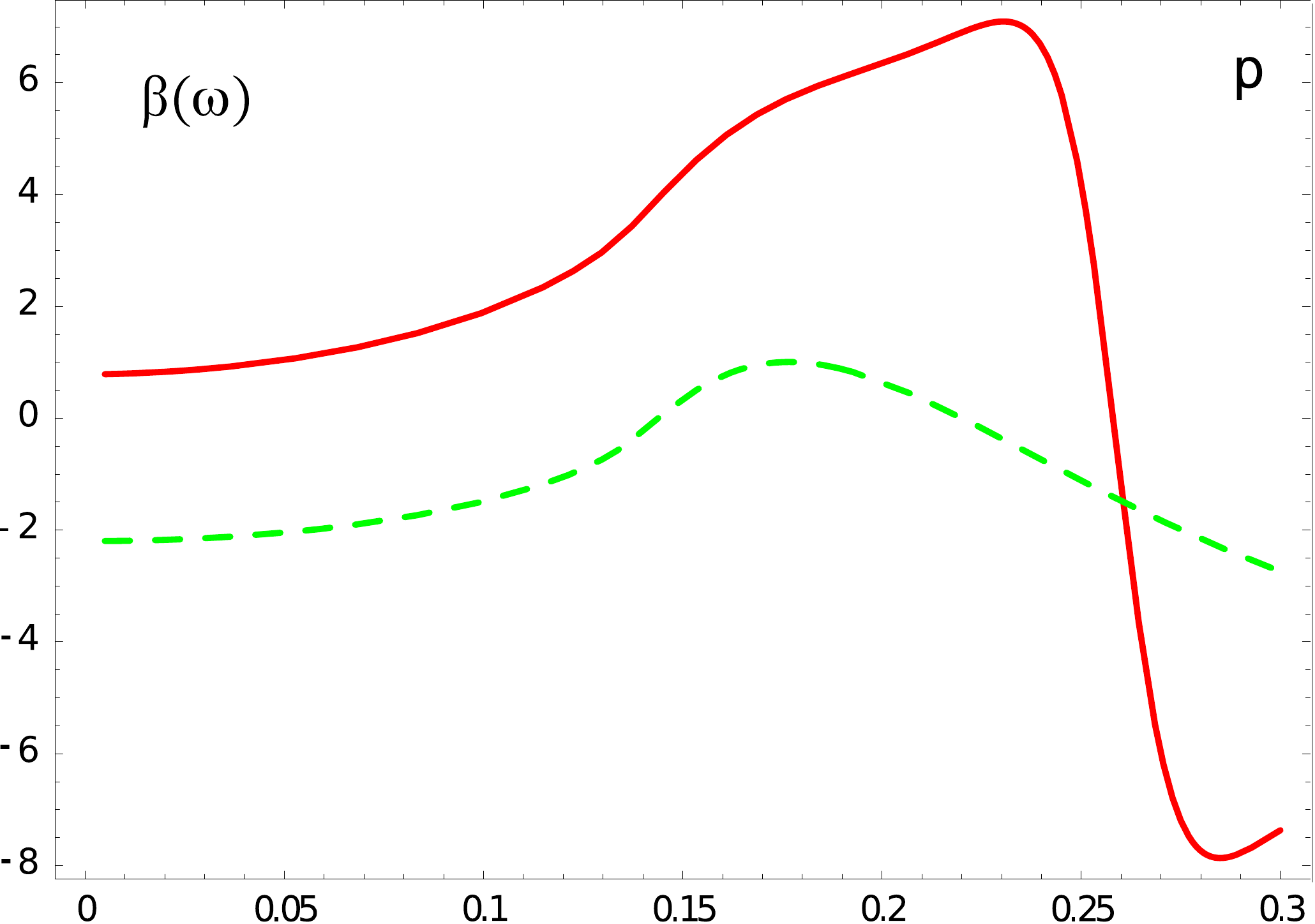}\tabularnewline
 & \tabularnewline
\end{tabular}
\par\end{centering}

\centering{}\caption{Dependencies of the proton electric and magnetic polarizabilities
(in $10^{-4}\,(fm^{3})$) on photon energy, $\omega$ (GeV), in the center-of-mass
reference frame. Green-dashed curves correspond to the meson-nucleon
loops contribution only; solid-red curves include the $\Delta$-resonance
pole contribution.}
\label{ff1}
\end{figure*}

The polarizabilities calculated for the proton with the photon energies
up to 300 MeV are shown in Fig.\ref{ff1}. It is evident that
below 50 MeV these polarizabilities have small energy dependence. For the neutron,
the energy dependencies of dynamical polarizabilities have very similar
behavior except the values are bigger on the absolute scale, so we
will only describe the dynamical polarizabilities for the proton. 

\begin{figure*}[!htpb]
\begin{centering}
\begin{tabular}{ccc}
\includegraphics[scale=0.25]{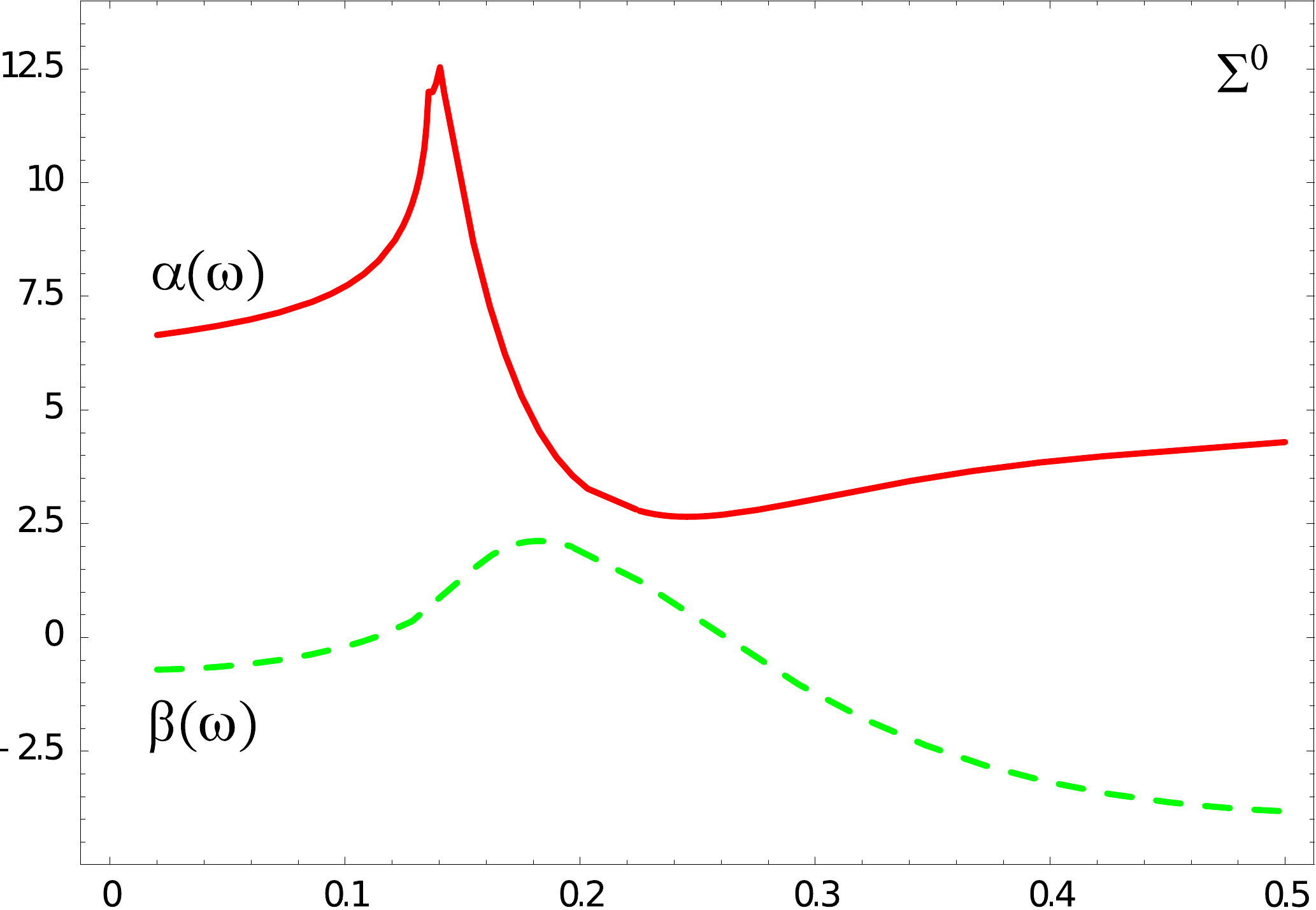} & \includegraphics[scale=0.25]{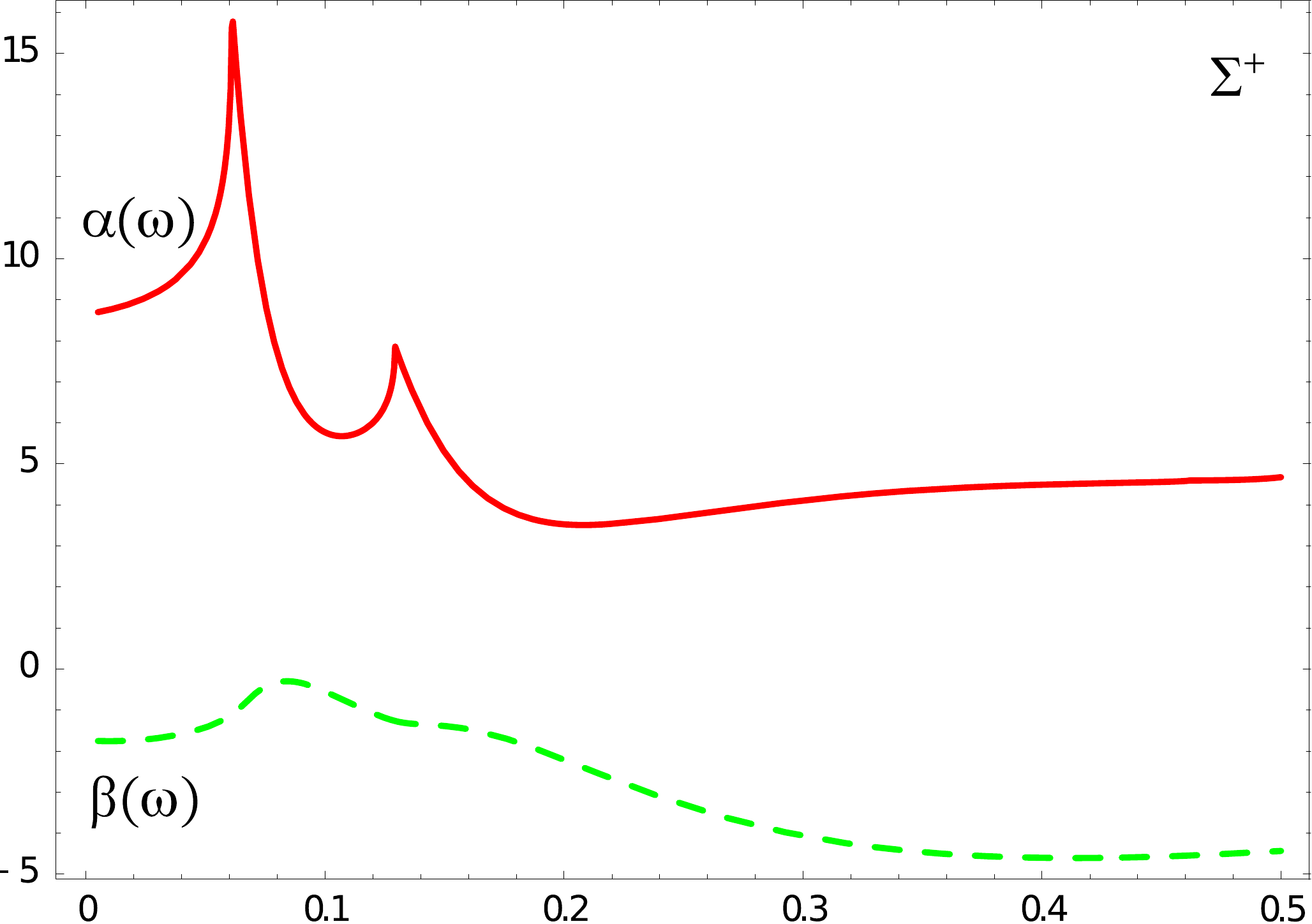} & \includegraphics[scale=0.25]{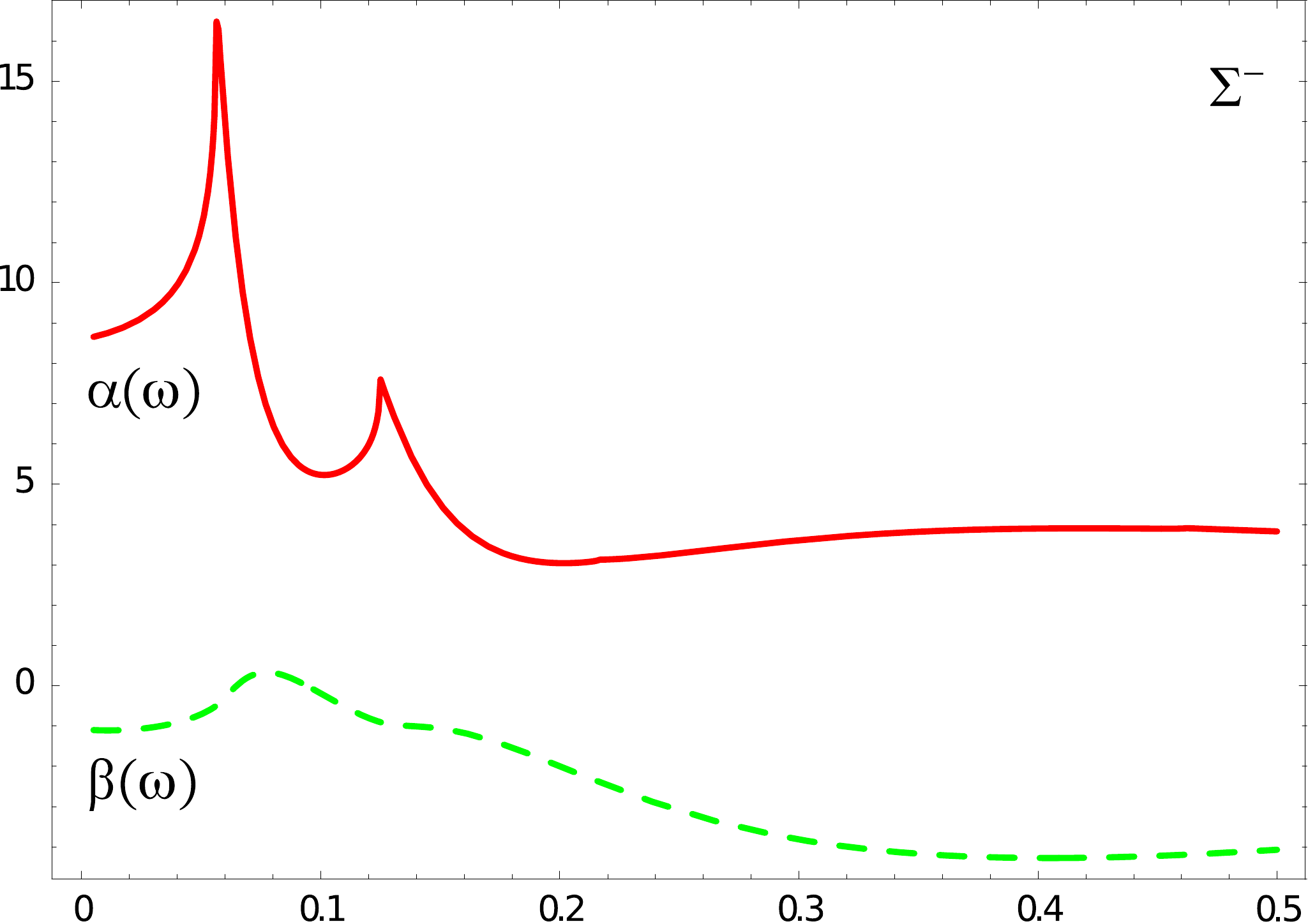}\tabularnewline
\includegraphics[scale=0.25]{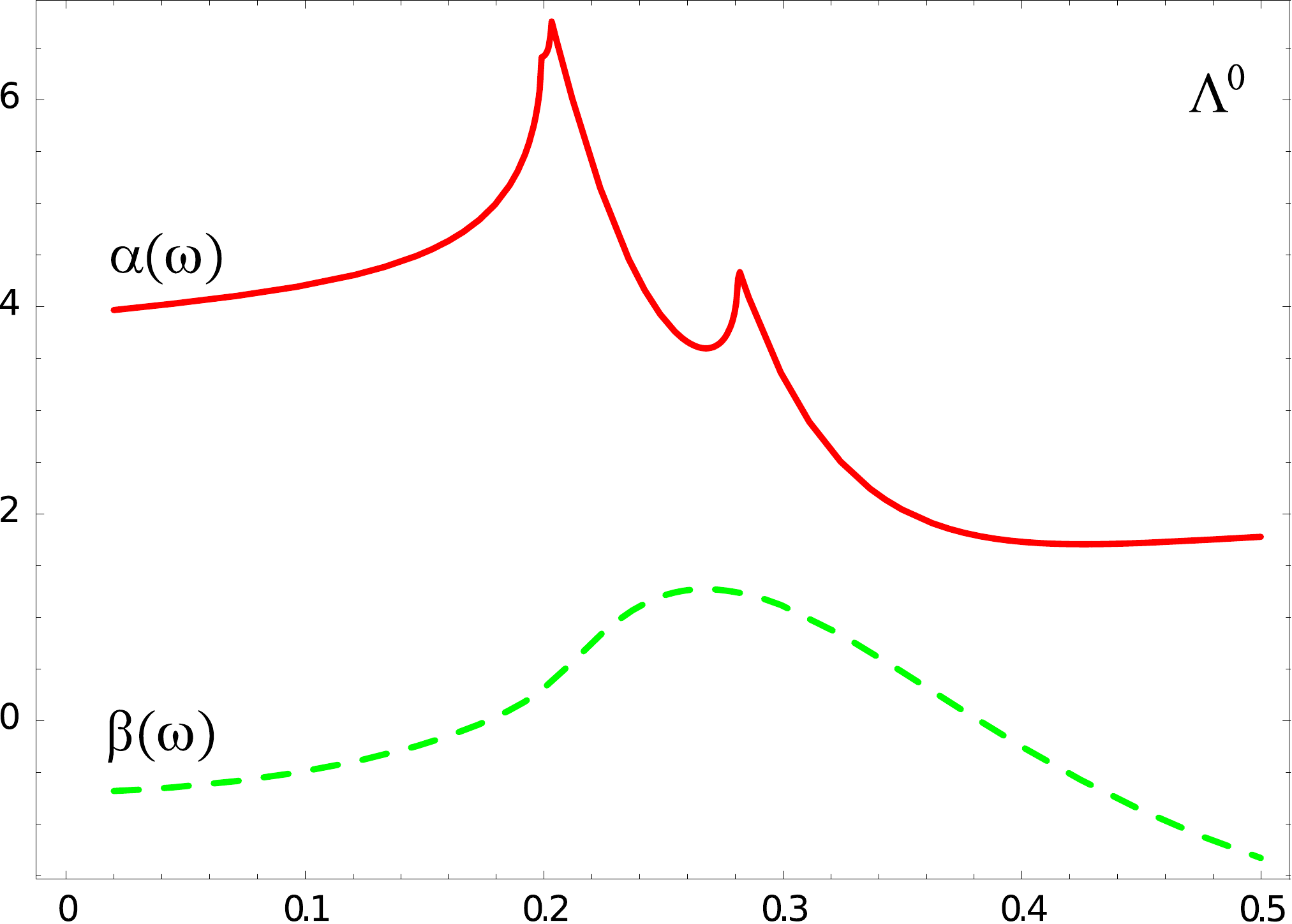} & \includegraphics[scale=0.25]{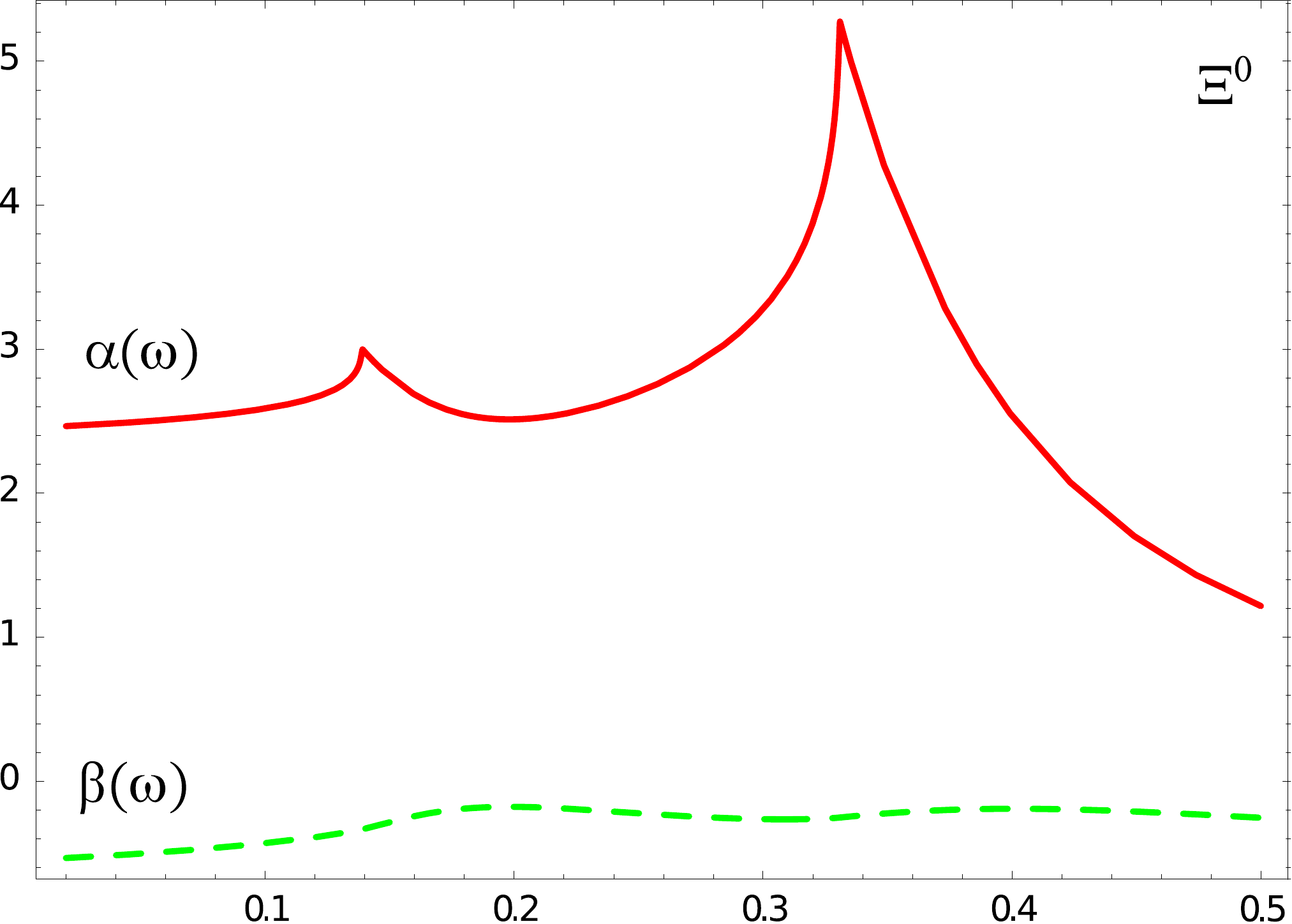} & \includegraphics[scale=0.25]{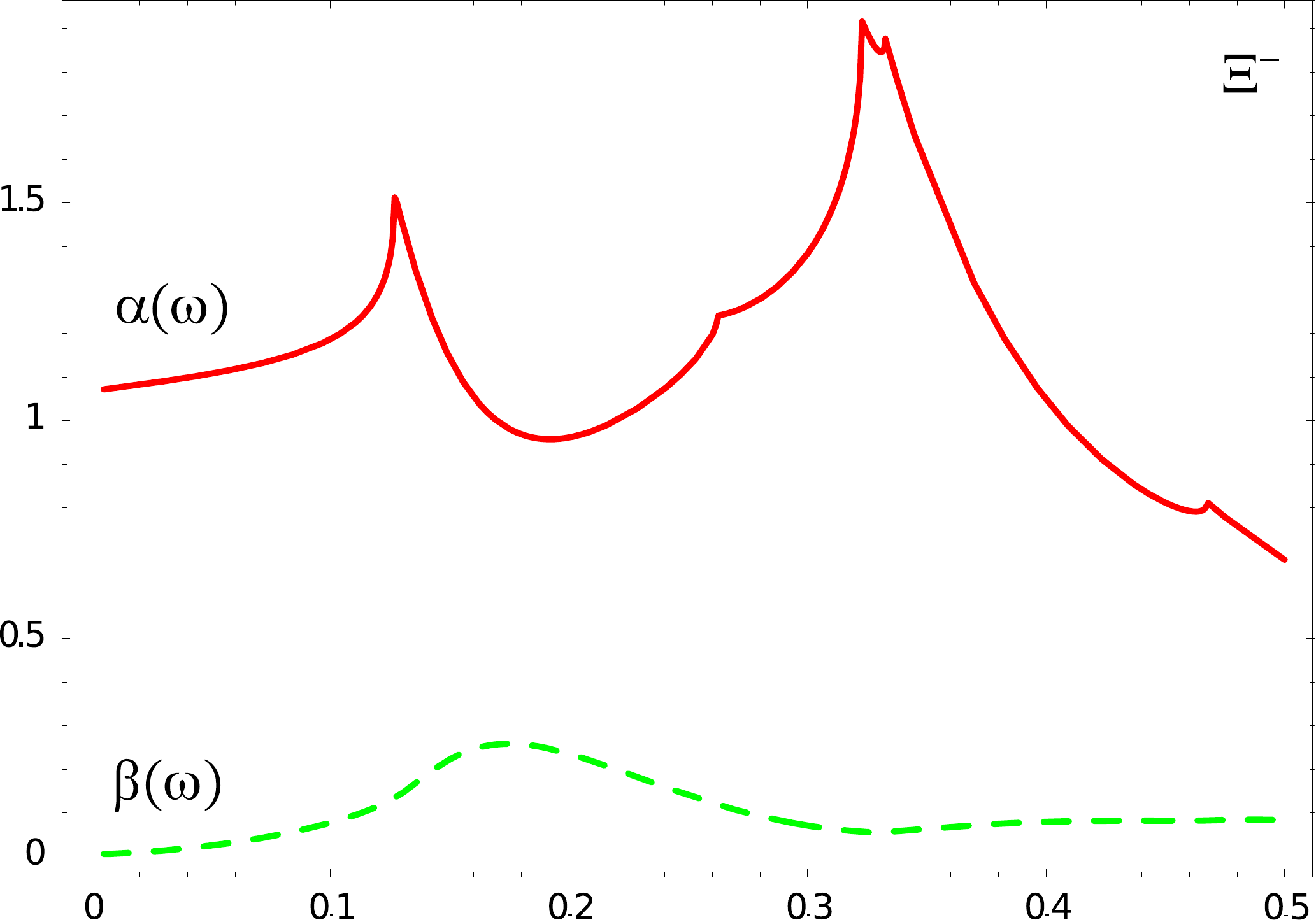}\tabularnewline
\end{tabular}
\par\end{centering}

\caption{Electric and magnetic dynamical polarizabilities of hyperons in units
of $10^{-4}\,(fm^{3})$ as a function of the photon energy, $\omega$ (GeV).
Here, the solid-red line represents the electric polarizability, and the dashed-green
line is the magnetic polarizability.}

\label{ff2}
\end{figure*}

The electric polarizability of the proton has very strong resonance
type dependence near the pion production threshold. The $\Delta$-pole
contribution has a small effect while consistently reducing $\alpha_{p}(\omega)$
values for all energies. Of course, to make final predictions for
the CHPT values of polarizabilities, we need to add the contribution from
the resonances in the loops of Compton scattering. Hence, in order
to compare our results with the experimental values, we have borrowed
the resonance loops results from the small-scale expansion (SSE) approach
\cite{SSE}. If no $\Delta$-pole contribution is added, the magnetic
polarizability in Fig.\ref{ff1} stays negative (diamagnetic) for
almost all the energies. The $\Delta$-pole contribution is very big
and shifts $\beta_{p}(\omega)$ from negative to positive (paramagnetic)
values for energies up to 250 MeV. This behavior is quite natural
since the pion loop calculations reflect magnetic polarizability coming
from the virtual diamagnetic pion cloud and the $\Delta$ resonance
contribution to $\beta_{p}(\omega)$ is driven by the strong paramagnetic
core of the nucleon. Our results for the proton polarizabilities calculated
in relativistic CHPT up to one-loop order including the $\Delta$-
pole and SSE contribution are the following (in units of $10^{-4}(fm^{3}))$:
\begin{eqnarray*}
 & \alpha_{p}=(7.38\,(\pi-\mbox{loop})-0.95\,(\Delta-\mbox{pole})+\nonumber \\
 &4.2\,(\mbox{SSE}))=10.63;\\ \\
 & \beta_{p}=(-2.20\,(\pi-\mbox{loop})+3.0\,(\Delta-\mbox{pole})+\nonumber \\
&0.7\,(\mbox{SSE}))=1.49.
\end{eqnarray*}
The static electric and magnetic polarizabilities for hyperons have
been calculated first in \cite{Meissner} in the heavy baryon chiral perturbation
theory. The dynamical electric and magnetic polarizabilities for hyperons
were first calculated in \cite{AB}. In Fig.\ref{ff2}, we provide
our results for dynamical electric and magnetic polarizabilities for
hyperons using the basis from Eq.(\ref{eq:4-1}) in the Compton scattering
amplitude. 

For all polarizabilities listed in Fig.(\ref{ff2}), the electric
polarizabilities have very similar resonant-type behavior near meson-production
thresholds and the magnetic polarizabilities for all hyperons have
negative low energy (static) values. Once again, it is important to
include both pole and loop resonance contributions for a complete
analysis. It is clear that for the all dynamical polarizabilities
of the SU(3) octet of baryons, the values are strongly governed by the
excitation mechanism reflected in the meson production peaks. Hence,
the study of these polarizabilities directly probes the internal degrees
of freedom governing the baryon structure at low energies.

\section{Conclusions}

In this work, we have evaluated the pion form factor and have investigated its influence on the behavior of the two-photon box pion electroproduction correction. We also calculated
the dynamical polarizabilities of mesons including the entire SU(3) octet
of mesons in the loop integrals using relativistic CHPT implemented
in CHM. The dynamical electric and magnetic polarizabilities
of the SU(3) octet of baryons were investigated in detail. 
We found that predictions of the chiral theory derived from
our calculations (up to one-loop order and not including resonances
in loop calculations) are somewhat consistent with the experimental
results. The dependencies for the range of photon energies covering
the majority of the meson photo production channels were analyzed.
These extensive calculations are made possible by the recent implementation
of semi-automatized calculations in CHPT, which
allows the evaluation of polarizabilities from Compton scattering up to next-to-the-leading
order. Our current goal is the calculation of dynamical polarizabilities with baryon
resonances in the loops. There is still some disagreement between heavy-baryon and relativistic versions of CHPT introducing theoretical
uncertainty. Clearly, further experimental work is needed
especially for the hyperon and strange mesons polarizabilities, which would help with the further development of the theory.

\section{Acknowledgements}
This work has been supported by the Natural Sciences and Engineering Research Council of Canada (NSERC).  

\nocite{*}
\bibliographystyle{elsarticle-num}







\end{document}